\documentclass[showpacs,10pt,twocolumn,prb]{revtex4}
\usepackage{amsmath}
\usepackage{amssymb}
\usepackage{graphics}
\usepackage{epsfig}
\usepackage{color}

\begin{document}

\title{Upper critical fields and two-band superconductivity in Sr$_{1-x}$Eu$%
_{x}$(Fe$_{0.89}$Co$_{0.11}$)$_{2}$As$_{2}$ ($x=0.203$ and $0.463$)}
\author{Rongwei Hu$^{1,\ast }$, Eun Deok Mun$^{2}$, M. M. Altarawneh$%
^{2,\dag }$, C. H. Mielke$^{2}$, V. S. Zapf$^{2}$, S. L. Bud'ko$^{1}$, P. C.
Canfield$^{1}$}
\affiliation{$^{1}$Ames Laboratory, U.S. DOE and Department of Physics and Astronomy,
Iowa State University, Ames, IA 50011, USA}
\affiliation{$^{2}$National High Magnetic Field Laboratory, Los Alamos National
Laboratory, Los Alamos, New Mexico 87545, USA}
\date{\today }

\begin{abstract}
The upper critical fields, $H_{c2}$ of single crystals of Sr$_{1-x}$Eu$_{x}$%
(Fe$_{0.89}$Co$_{0.11}$)$_{2}$As$_{2}$ ($x=0.203$ and $0.463$) were
determined by radio frequency penetration depth measurements in pulsed
magnetic fields. $H_{c2}$ approaches the Pauli limiting field but shows an
upward curvature with an enhancement from the orbital limited field as
inferred from Werthamer-Helfand-Hohenberg theory. We discuss the temperature
dependence of the upper critical fields and the decreasing anisotropy using
a two-band BCS model.
\end{abstract}

\pacs{74.25.Dw, 74.25.Op, 74.70.Dd}
\maketitle

The upper critical fields $H_{c2}$ and its anisotropy are fundamental
characteristics of a type-II superconductor, they provide information about
the underlying electronic structure and can shed light on\ the mechanism of
Copper pair breaking. Therefore for both understanding of superconductivity
and potential application, extensive studies of $H_{c2}$ have been performed
on the recently discovered FeAs-based superconductors. Large upper critical
fields have been observed for FeAs superconductors.\cite{Shahbazi}$^{-}$\cite%
{NNi} More interestingly, they exhibit pronounced upward curvature of $%
H_{c2} $, implying multiband nature of these materials.\cite{Kano}$^{,}$\cite%
{Hunte}$^{-}$\cite{Baily} In contrast to the high $T_{c}$ cuprates with very
large anisotropy, although they both possess a layered crystal structure,
measurements of $H_{c2}$ of the FeAs superconductors have revealed that the
anisotropic ratio $\gamma =H_{c2}^{ab}/H_{c2}^{c}$ decreases with decreasing
temperature and becomes nearly isotropic at low temperatures for the 122 and
111 type of FeAs materials.\cite{ZhangJL}$^{-}$\cite{Hunte}$^{,}$\cite%
{YuanHQ}

Previous study of the Eu doped Sr(Fe$_{0.88}$Co$_{0.12}$)$_{2}$As$_{2}$
demonstrated the interaction between the FeAs-based superconductivity and
magnetism due to Eu$^{2+}$: in the disordered paramagnetic region of Eu$%
^{2+} $, superconductivity is weakly suppressed by spin-flip scattering off
the local magnetic moments of Eu$^{2+}$; it is further suppressed with
developing long range antiferromagnetic order of Eu$^{2+}$ and coexists with
antiferromagnetism of Eu$^{2+}$ as long as $T_{c}>T_{N}$.\cite{Rongwei} It
is of great interest to see how the superconductivity is modified by the
magnetism of Eu$^{2+}$ by mapping out the $H-T$ phase diagram.

Moreover, in the study of the interplay of superconductivity and magnetism,
it is proposed by Jaccarino \textit{et al}\cite{Jaccarino} that for certain
rare earth ferromagnetic metal, the external magnetic field, which in
general inhibits superconductivity, may be cancelled by the effective
exchange field $H_{eff}$ of the magnetic moments, imposed on the conduction
electrons, when $H_{eff}$ is opposite to the direction of applied field.
Therefore superconductivity can occur in two domains, one at low field,
where pair-breaking field is still small, and one at high field in the
compensation region. Experimentally, an anomalous enhancement of $H_{c2}$
was first reported by Fischer \textit{et al}\cite{Fischer}\textit{\ }in Sn$%
_{1.2(1-x)}$Eu$_{x}$Mo$_{6.35}$S$_{8}$ and Pb$_{1-x}$Eu$_{x}$Mo$_{6.35}$S$%
_{8}$ chevrel phases and was suggested to be related to Jaccarino-Peter
effect. A magnetic field induced superconductivity in the $H_{c2}-T$ phase
diagram was indeed observed in Eu$_{0.75}$Sn$_{0.25}$Mo$_{6}$S$_{7.2}$Se$%
_{0.8}$ and fitted well with the Jaccarino-Peter scenario.\cite{Meul}
Therefore, the properties of Sr$_{1-x}$Eu$_{x}$(Fe$_{0.89}$Co$_{0.11}$)$_{2}$%
As$_{2}$, as possible candidates for observation of Jaccarino-Peter effect,
are worth investigating.

In this paper we report the upper critical fields of Sr$_{1-x}$Eu$_{x}$(Fe$%
_{0.89}$Co$_{0.11}$)$_{2}$As$_{2}$ ($x=0.203$ and $0.463$) single crystals
determined by radio frequency contactless penetration depth measurements.
The two selected samples are the representative concentrations in the
disordered paramagnetic region and coexistence region of superconductivity
and antiferromagnetism. We find that for both concentrations the curves of $%
H_{c2}(T)$ can be consistently explained by the two-band model and the
anisotropy decreases with temperature approaching an isotropic state at low
temperatures.

Single crystals of Sr$_{1-x}$Eu$_{x}$(Fe$_{0.89}$Co$_{0.11}$)$_{2}$As$_{2}$
were grown from self flux, as describe in Ref. 12. Chemical composition was
determined by wavelength dispersive x-ray spectroscopy (WDS) in a JEOL
JXA-8200 electron microscope. Magnetic susceptibility was measured in a
Quantum Design MPMS. The temperature and magnetic field dependences of the
electrical resistance were measured using the four probe ac ($f$ = 16Hz)
technique in a Quantum Design PPMS. Radio frequency (rf) contactless
penetration depth measurements were performed on the single-crystal sample
in a 60 T pulsed field magnet with a 10 ms rise time and a 40 ms extended
decay. The rf technique is highly sensitive to small changes~($\sim $1--5
nm)~ in the rf penetration depth, thus it is an accurate method for
determining the upper critical field in anisotropic superconductors.\cite%
{Mielke} Small single crystals were selected because of the eddy current
heating in pulsed field. To determine the upper critical-field anisotropy,
the single crystal was measured in two $H\parallel ab$ and $H\parallel c$
configurations. More details about this technique can be found in Ref.4, 17,
18.

\begin{figure}[tbp]
\centerline{\includegraphics[scale=0.35]{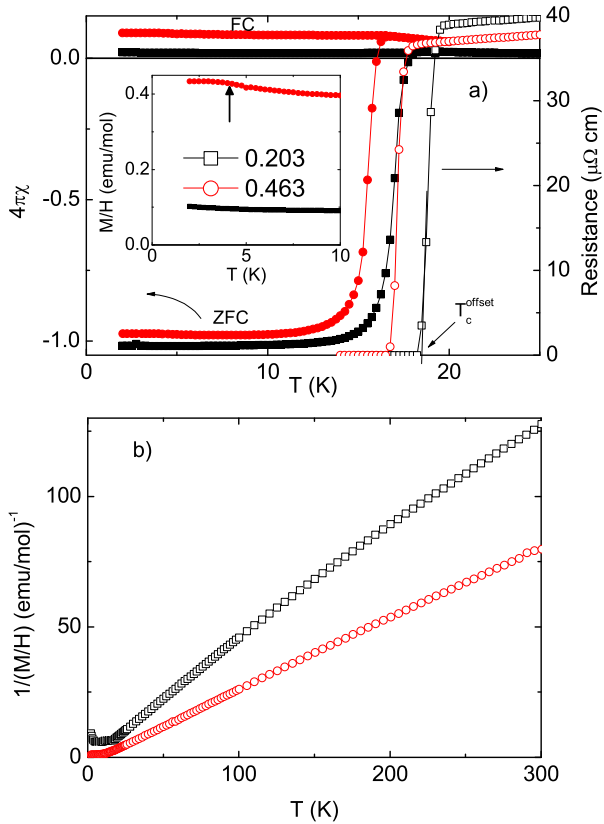}} \vspace*{-0.3cm}
\caption{a) Low temperature magnetic susceptibility measured in a magnetic
field 100 Oe applied in \textit{ab} plane and resistivity in zero field.
Inset shows an zoom-in view of the field-cooled curve, arrow indicates the
antiferromagnetic transition. b) Inverse in-plane magnetic susceptibility
measured in 10 kOe.}
\end{figure}

\begin{figure*}[tbp]
\centerline{\includegraphics[scale=0.6]{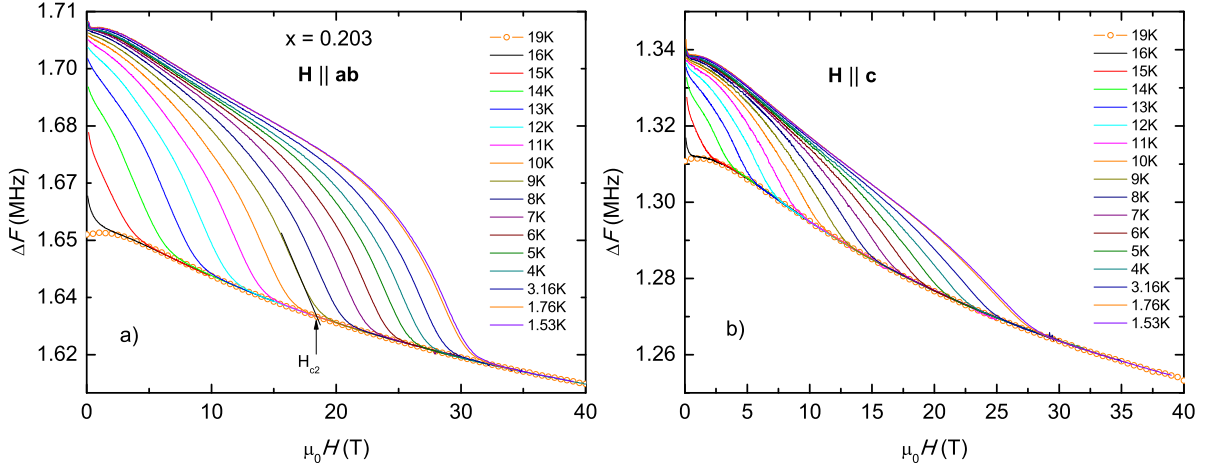}} \vspace*{-0.3cm}
\caption{Frequency shift ($\Delta F$) as a function of magnetic field for $%
H\parallel ab$ and $H\parallel c$ for Eu20 sample at selected temperatures.
Open symbols are $\Delta F$ taken at 19 K as a normal state, background
signal. It shows the criterion to determine $H_{c2}$. }
\end{figure*}

The actual compositions of the two samples determined by WDS were Sr$%
_{0.797} $Eu$_{0.203}$(Fe$_{0.888}$Co$_{0.112}$)$_{2}$As$_{2}$ and Sr$%
_{0.537}$Eu$_{0.463}$(Fe$_{0.885}$Co$_{0.115}$)$_{2}$As$_{2}$. For brevity,
we denote them as Eu20 and Eu46 sample in the following text. The Co
concentrations are consistent with the optimal doping, $x\sim 0.12$, for
Sr(Fe$_{1-x}$Co$_{x}$)$_{2}$As$_{2}$ as in Ref. 12. Figure 1(a) shows the
low temperature magnetic susceptibility and resistivity of the two samples.\
The large diamagnetic shielding indicates bulk superconductivity. The
superconducting transition temperatures inferred from the first deviation
point from the normal magnetic susceptibility of the zero-field-cool curve
are 18 K and 16.2 K for Eu20 and Eu46 respectively. The Eu46 sample shows a
weak anomaly due to antiferromagnetic ordering of Eu$^{2+}$ at 3.5 K as
indicated in the inset of Fig. 1(a). The $T_{c}$ in resistivity as inferred
from by extrapolating the steepest slop to zero resistance are 18.3 K and
16.8 K for the two samples, in agreement with the magnetic susceptibility
measurements. The inverse in-plane magnetic susceptibility measured in 10
kOe of the two samples is plotted in Fig.1 (b). The Curie-Weiss fits above
150 K give an estimated Eu concentration of $0.215$ and $0.469$ by assuming
7.94 $\mu _{B}/ $Eu$^{2+}$ ion. Thus all the above observations are
consistent with those in Ref. 12 and show that Eu20 is in the disordered
paramagnetic region of Eu$^{2+}$ and Eu46 is in the coexistence region of
superconductivity and antiferromagnetism.

\begin{figure*}[tbp]
\centerline{\includegraphics[scale=0.6]{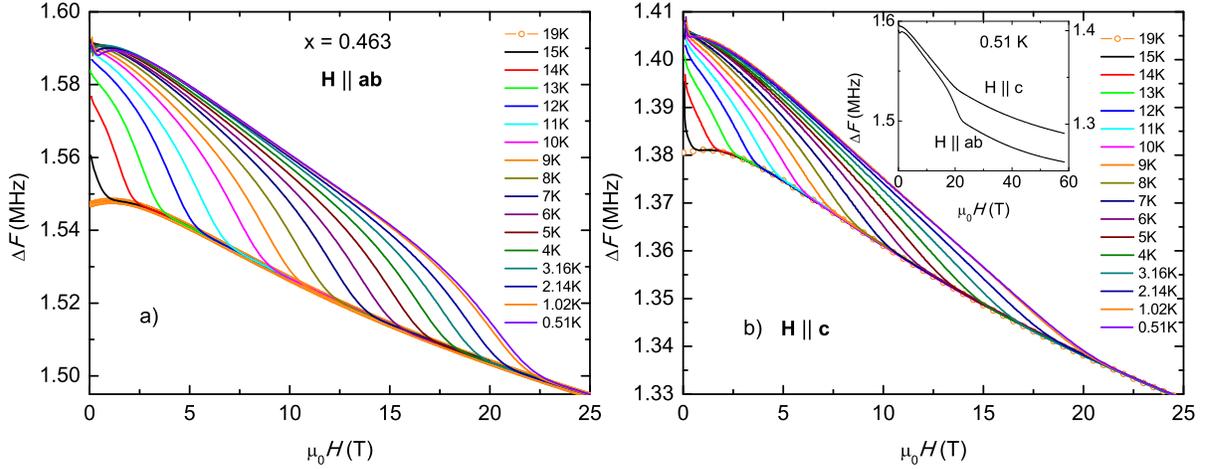}} \vspace*{-0.3cm}
\caption{$\Delta F$ as a function of magnetic field for $H\parallel ab$ and $%
H\parallel c$ for Eu46. Inset in (b) shows the measurements up to 60 T at
the base temperature of 0.51 K.}
\end{figure*}

The frequency shift as a function of magnetic field applied parallel and
perpendicular to the \textit{ab }plane at different temperatures from 1.5 to
19 K for Eu20 is shown in Fig. 2. The normal state has a smooth and nearly
linear field dependence as manifested by the 19 K curve.\cite{Eun} $H_{c2}$
is identified as the point at which the slope of the $\Delta F$ intercepts
the normal state background of 19 K. Other criterion, e.g. first point
deviating from the normal state background can be used and the difference
between these two criteria is taken as the error bar for $H_{c2}$. For $%
H\parallel c$ in Fig. 2(b), the sample has a weaker coupling to the
detection coil, resulting in a smaller but still easily resolvable frequency
shift. The same rf measurements were performed on Eu46 sample for both
orientations for temperatures down to 0.51 K and shown in Fig. 3. In the
previous study in Ref. 12, it has been shown that the Eu$^{2+}$ moments
undergo a metamagnetic transition from antiferromagnetic to ferromagnetic
above a magnetic field of 4 kOe. Thus it behaves as a superconductor with
ferromagnetically coupled Eu$^{2+}$ moments at low temperature high field.
In order to look for possible Jaccarino-Peter effect, the frequency shift of
Eu46 sample was measured in field up to 60 T at the base temperature 0.51 K
for both directions (inset in Fig. 3(b)). No anomaly associated with
superconductivity can be observed in high fields. So either the magnetic
field is still too low to compensate the exchange field or the exchange
field has the same sign as the external field then no cancellation can be
realized.

\begin{figure}[tbp]
\centerline{\includegraphics[scale=0.35]{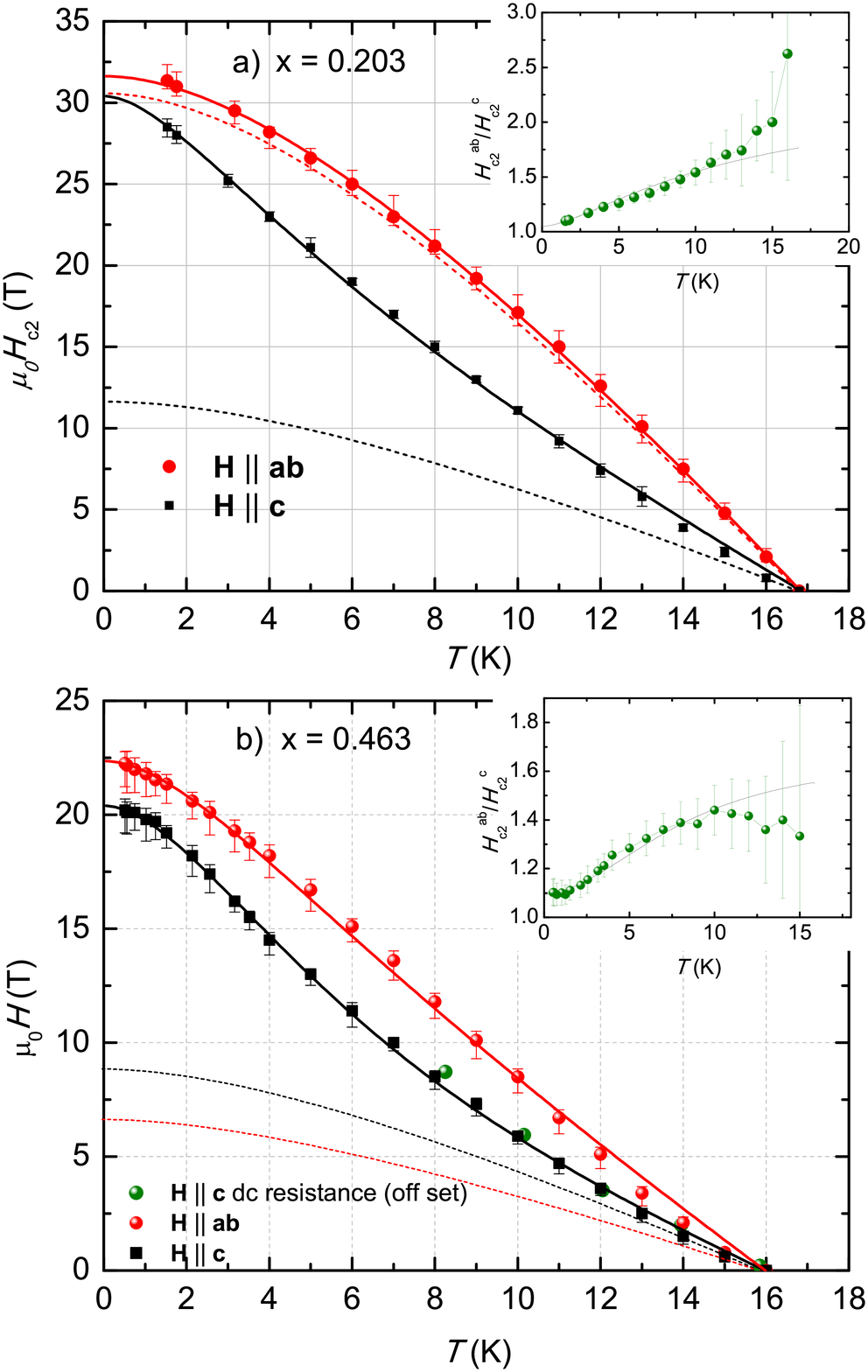}} \vspace*{-0.3cm}
\caption{Anisotropic $H_{c2}(T)$ for Eu20 and Eu46 single crystals. The
green circles in (b) are obtained from the resistivity measurement, in
excellent agreement with the pulsed field rf shift measurement. The dotted
lines are fits to WHH formula. The solid lines are fits to the two-band
model. Insets shows the temperature dependence of the anisotropy $\protect%
\gamma =H_{c2}^{ab}/H_{c2}^{c}$ and the solid lines are the calculated curve
of the two-band model fits.}
\end{figure}

Figure 4 shows the $H_{c2}(T)$ curves for $H\parallel ab$ ($H_{c2}^{ab}$)
and $H\parallel c$ ($H_{c2}^{c}$) of both samples. For the Eu20 sample, $%
H_{c2}^{ab}$ is almost linear close to $T_{c}$, a traditional
Werthamer-Helfand-Hohenberg (WHH) behavior, but $H_{c2}^{c}$ exhibits a
significant upward curvature. This negative curvature is even more
pronounced for the Eu46 sample in Fig. 4(b) for both field orientations. The
dashed lines in Fig. 4 are fits to the conventional one-band WHH theory.\cite%
{WHH} The $H_{c2}$ values from direct measurements are far above the
prediction of WHH theory, except for the $H\parallel ab$ curve of Eu20
sample (see later discussion). The other mechanism for limiting $H_{c2}$ is
the Pauli spin paramagnetic effect as a result of Zeeman effect exceeding
the condensation energy of Copper pairs, given by $\mu _{0}H_{p}=$ $%
1.84T_{c} $ for isotropic s-wave pairing.\cite{Clogston} $\mu _{0}H_{p}$ is
estimated to be 30.9 T and 29.4 T for Eu20 and Eu46 respectively. These
values are close to the experimental results extrapolated to 0 K, implying
that the Pauli paramagnetic effect might be the dominant pair breaking
mechanism for limiting the upper critical fields in these compounds. 
\begin{table*}[tbp]
\caption{Parameters of the fits to the two-band model for Sr$_{x}$Eu$_{x}$(Fe%
$_{0.89}$Co$_{0.11}$)$_{2}$As$_{2}$}%
\begin{tabular}{ccccccc}
\hline\hline
$x$ & $\left\langle 
\begin{array}{cc}
D_{1}^{ab} & D_{1}^{c} \\ 
D_{2}^{ab} & D_{2}^{c}%
\end{array}%
\right\rangle (cm^{2}/s)$ & $\left\langle 
\begin{array}{cc}
\lambda _{11} & \lambda _{12} \\ 
\lambda _{21} & \lambda _{22}%
\end{array}%
\right\rangle $ & 
\begin{tabular}{l}
$\mu _{0}H_{c2}^{ab}(0)$ \\ 
$(T)$%
\end{tabular}
& 
\begin{tabular}{l}
$\mu _{0}H_{c2}^{c}(0)$ \\ 
$(T)$%
\end{tabular}
& 
\begin{tabular}{l}
$\xi ^{ab}(0)$ \\ 
$(nm)$%
\end{tabular}
& 
\begin{tabular}{l}
$\xi ^{c}(0)$ \\ 
$(nm)$%
\end{tabular}
\\ \hline
$0.203$ & $\left\langle 
\begin{array}{cc}
0.16 & 1.35 \\ 
0.36 & 0.15%
\end{array}%
\right\rangle $ & $\left\langle 
\begin{array}{cc}
0.19 & 0.194 \\ 
0.194 & 0.21%
\end{array}%
\right\rangle $ & $31.6$ & $30.4$ & $3.3$ & $3.2$ \\ \hline
$0.463$ & $\left\langle 
\begin{array}{cc}
0.79 & 2.27 \\ 
0.28 & 0.25%
\end{array}%
\right\rangle $ & $\left\langle 
\begin{array}{cc}
0.2 & 0.082 \\ 
0.082 & 0.2%
\end{array}%
\right\rangle $ & $22.4$ & $20.4$ & $4.0$ & $3.7$ \\ \hline\hline
\end{tabular}%
\end{table*}

On the other hand, the anomalous upward curvature of $H_{c2}(T)$ has been
observed in other multiband systems like MgB$_{2}$\cite{Gurevich} and
recently in FeAs superconductors e.g. Ba(Fe$_{0.9}$Co$_{0.1}$)$_{2}$As$_{2}$%
\cite{Kano}, LaFeAsO$_{0.89}$F$_{0.11}$\cite{Hunte}, NdFeAsO$_{0.7}$F$_{0.3}$%
\cite{Jaroszynski} and Sr(Fe$_{0.9}$Co$_{0.1}$)$_{2}$As$_{2}$ thin film\cite%
{Baily} and explained within a two-band BCS model by taking into account the
inter and intra band scattering in $H_{c2}$.\cite{Gurevich} In the two-band
s-wave theory, the intra and interband interaction is described by a $%
2\times 2$ matrix of the BCS coupling constants $\lambda _{mn}$, for which $%
\lambda _{11}$ and $\lambda _{22}$ quantify the intraband coupling and\ $%
\lambda _{12}$ and $\lambda _{21}$ describe interband coupling. $H_{c2}$ is
described by a parametric equation\cite{Gurevich}%
\begin{eqnarray*}
\ln \frac{T}{T_{c0}} &=&-(U(h)+U(\frac{D_{2}}{D_{1}}h)+\frac{\lambda _{0}}{w}%
)/2 \\
&&+[(U(h)-U(\frac{D_{2}}{D_{1}}h)-\frac{\lambda _{-}}{w})^{2}/4+\frac{%
\lambda _{12}\lambda _{21}}{w}]^{1/2} \\
U(h) &=&\psi (1/2+h)-\psi (1/2) \\
H_{c2} &=&2\phi _{0}k_{B}Th/\hbar D_{1}
\end{eqnarray*}

where $\psi (x)$ is the digamma function, $\phi _{0}$ is the flux quantum, $%
k_{B}$ is the Boltzmann constant, $\hbar $ is the plank constant, $D_{1,2}$
are the anisotropic diffusivities of each band, for $H_{c2}^{ab}$ the
diffusivity $D_{1}$ should be replaced by $(D_{1}^{ab}D_{1}^{c})^{1/2}$, $%
\lambda _{-}=\lambda _{11}-\lambda _{22}$, $\lambda _{0}=(\lambda
_{-}^{2}+4\lambda _{12}\lambda _{21})^{1/2}$, $w=\lambda _{11}\lambda
_{22}-\lambda _{12}\lambda _{21}$. Since only the product of $\lambda _{12}$
and $\lambda _{21}$ appears in the equation, we can assume $\lambda
_{12}=\lambda _{21}$. The fits to both $H_{c2}^{ab}$ and $H_{c2}^{c}$ for
each sample are performed simultaneously in a self-consistent manner. The
model fits the data remarkably well, it captures the main features of the $%
H_{c2}$ curves. The fitting parameters are listed in Table I. In terms of
diffusivity, the two bands exhibit strong asymmetry, i.e. the diffusivity
ratio $\sqrt{D_{2}^{ab}D_{2}^{c}}/\sqrt{D_{1}^{ab}D_{1}^{c}}\sim 0.5$ and $%
0.2$ for Eu20 and Eu46 respectively. Thus superconductivity results from an
anisotropic band with high diffusivity and a more isotropic band with
smaller diffusivity. It should be noted that for the Eu20 sample $H_{c2}^{c}$
shows negative curvature whereas $H_{c2}^{ab}$ shows behavior similar to
that conforms with the conventional WHH theory. The two types of curvature
for different field orientations have also been observed in Ba(Fe$_{0.93}$Co$%
_{0.07}$)$_{2}$As$_{2}$.\cite{Gasparov} But here we are describing both of
them within the two-band model. For equal diffusivities of the two bands,
i.e. $\eta =D_{2}/D_{1}=1$, the parametric equation of above reduces to the
one-gap de-Gennes-Maki formula in WHH theory, $\ln t+U(h)=0$.\cite{Gurevich}
The diffusivity ratio of the Eu20 sample, $\eta
^{ab}=D_{2}^{ab}/(D_{1}^{ab}D_{1}^{c})^{1/2}$ and $\eta
^{c}=D_{2}^{c}/D_{1}^{c}$, is 0.77 and 0.11 for $H\parallel ab$ and $%
H\parallel c$ respectively. Therefore it is reasonable to expect $%
H_{c2}^{ab} $ with near unity $\eta $ to show WHH-like behavior in contrast
to $H_{c2}^{c}$ with much lower $\eta $ to be two-band-like.

The Eu20 sample shows strong interband pairing, i.e. $\lambda _{12}\lambda
_{21}\simeq \lambda _{11}\lambda _{22}$, whereas the two bands become more
non-interacting in the Eu46 sample, as indicated by $\lambda _{12}\lambda
_{21}\ll \lambda _{11}\lambda _{22}$. It is noteworthy that the intraband
pairing strength, $\lambda _{11}$ and $\lambda _{22}$, remains almost
unchanged for Eu concentration increases from $0.203$ to $0.463$, only the
interband coupling decreases, with $T_{c}$ decreases slowly from 16.8 K to
16 K. This observation may imply that superconductivity could be dominated
by the intraband pairing and not particularly sensitive to disorder and
interband scattering. With the fitted values of $H_{c2}$ at 0 K, we can
estimate the anisotropic coherence length using $\xi ^{ab}=\sqrt{\phi
_{0}/2\pi H_{c2}^{c}}$ and $\xi ^{c}=\phi _{0}/2\pi \xi ^{ab}H_{c2}^{ab}$
(Table I). Both $\xi ^{ab}$ and $\xi ^{c}$ are much larger than the spacing
between the superconducting FeAs layers ($\sim 6$\AA ) in Sr$_{1-x}$Eu$_{x}$%
(Fe$_{0.89}$Co$_{0.11}$)$_{2}$As$_{2}$, suggesting a 3D characteristic of
superconductivity.

The anisotropy of $H_{c2}$ is plotted in the insets of Fig. 4. Both $\gamma $
decrease with decreasing temperature and approach 1 at zero temperature. It
is qualitatively similar to that of the LiFeAs\cite{ZhangJL}, (Ba,K)Fe$_{2}$%
As$_{2}$\cite{Altarawneh}$^{,}$\cite{YuanHQ} and LaFeAsO$_{0.89}$F$_{0.11}$.%
\cite{Hunte} The isotropy of $H_{c2}$ in FeAs superconductors with different
carrier dopings is unexpected since distinctive hole and electron Fermi
surfaces may be responsible for superconductivity with different dopings.
For our Eu20 and Eu46 samples, there could be two factors contributing to
the decreasing anisotropy: i) at low temperature, band 2 with lower band
anisotropy $D_{2}^{ab}/D_{2}^{c}\sim 2.4-1.1$ may become more important than
band 1 with $D_{1}^{ab}/D_{1}^{c}\sim 0.12-0.35$; ii) the two bands have
opposing anisotropy of diffusivity, for band 1, $(D_{1}^{ab}/D_{1}^{c})<1$,
whereas for band 2, $(D_{2}^{ab}/D_{2}^{c})>1$. The calculated $\gamma $
from the fits are shown as the solid lines in the insets. They well
reproduce the temperature dependence of $\gamma $ within error bars.

To summarize, we measured the anisotropic $H_{c2}(T)$ for single crystals of
Sr$_{1-x}$Eu$_{x}$(Fe$_{0.89}$Co$_{0.11}$)$_{2}$As$_{2}$ ($x=0.203$ and $%
0.463$). Despite the presence of Eu$^{2+}$ moment, the Jaccarino-Peter
effect is not observed up to 60 T at base temperature of 0.5 K, it may be
intrinsically absent in this system or higher field is needed. $H_{c2}$
deviates from the WHH behavior as manifested by the upward curvature and is
probably limited by the Pauli paramagnetic pair breaking. The temperature
dependence of $H_{c2}$ is well described by a model of two bands with
opposing anisotropy and large diffusivity difference. The $H_{c2}$ becomes
more isotropic at low temperature.

This work was carried out at the Iowa State University and supported by the
AFOSR-MURI grant \#FA9550-09-1-0603 (R. H. and P. C. C.). Part of this work
was performed at Ames Laboratory, US DOE, under contract \# DE-AC02-07CH
11358 (S. L. B. and P. C. C.). S. L. B. was also partially supported by the
State of Iowa through the Iowa State University. Work at the NHMFL is
supported by the NSF, the DOE and the State of Florida.

*Present address: Center for Nanophysics \& Advanced Materials and
Department of Physics, University of Maryland, College Park MD 20742-4111,
USA.

\dag Present address: Department of physics, Mu'tah University, Mu'tah,
Karak, 61710, Jordan.

\end{document}